\documentclass[12pt]{article}
\usepackage{graphicx,amsfonts,amsmath,amssymb,amsthm,mathrsfs,url,hyperref}

\title{Hamiltonians Without Ultraviolet Divergence for Quantum Field Theories}

\author{
Stefan Teufel\footnote{Mathematisches Institut,
     Eberhard-Karls-Universit\"at, Auf der Morgenstelle 10, 72076
     T\"ubingen, Germany}\:\,\footnote{E-mail:
     stefan.teufel@uni-tuebingen.de}~~and
Roderich Tumulka$^*$\footnote{E-mail:
     roderich.tumulka@uni-tuebingen.de}
}
\date{February 3, 2020}

\newcommand{\Hilbert}{\mathscr{H}}
\newcommand{\Fock}{\mathscr{F}}
\newcommand{\conf}{\mathcal{Q}}
\newcommand{\Q}{\mathcal{Q}}

\renewcommand{\Im}{\mathrm{Im}}

\newcommand{\RRR}{\mathbb{R}}
\newcommand{\CCC}{\mathbb{C}}
\newcommand{\SSS}{\mathbb{S}}

\newcommand{\scp}[2]{\langle #1|#2 \rangle}
\newcommand{\Bscp}[2]{\Bigl\langle #1\Big|#2 \Bigr\rangle}
\newcommand{\Laplace}{\Delta} 

\newcommand{\orig}{{\mathrm{orig}}}
\newcommand{\cutoff}{{\mathrm{cutoff}}}
\newcommand{\inter}{{\mathrm{inter}}}

\newcommand{\vj}{\boldsymbol{j}}
\newcommand{\vk}{\boldsymbol{k}}
\newcommand{\vp}{\boldsymbol{p}}
\newcommand{\vq}{\boldsymbol{q}}

\newcommand{\vx}{\boldsymbol{x}}
\newcommand{\vy}{\boldsymbol{y}}

\newcommand{\vomega}{\boldsymbol{\omega}}
\newcommand{\vzero}{\boldsymbol{0}}

\newcommand{\domain}{\mathscr{D}}
\newtheorem{thm}{Theorem}
\newcommand{\be}{\begin{equation}}
\newcommand{\ee}{\end{equation}}

\begin{document}
\maketitle
\begin{abstract}
We propose a way of defining  Hamiltonians for quantum field theories without any renormalization procedure. The resulting Hamiltonians, called IBC Hamiltonians,  are mathematically well-defined (and in particular, ultraviolet finite) without an  ultraviolet cut-off such as smearing out the particles over a nonzero radius; rather, the particles are assigned radius zero. These Hamiltonians agree with those obtained through renormalization  whenever both are known to exist. We describe explicit examples of IBC Hamiltonians.  Their definition, which is best expressed in the particle--position representation of the wave function, involves a kind of boundary condition on the wave function, which we call an \emph{interior--boundary condition} (IBC). The relevant configuration space is one of a variable number of particles, and the relevant boundary consists of the configurations with two or more particles at the same location. The IBC relates the value (or derivative) of the wave function at a boundary point to the value of the wave function at an interior point (here, in a sector of configuration space corresponding to a lesser number of particles).

\medskip

\end{abstract}

\newpage
\tableofcontents

\section{Introduction}

In many quantum field theories (QFTs), the formulas that one obtains for the Hamiltonian (by means of quantization or other heuristics) contain terms for the creation and annihilation of particles that are ultraviolet (UV) divergent. 
The problem can be avoided by a UV cut-off, i.e., by discretizing space or treating 
the electron (and other particles) not as a point but smearing it out instead over a small positive radius; however, these procedures tend to break the Lorentz invariance, and there is no empirical evidence for either discrete space or a positive electron radius \cite{GJ85,GJ87}.
Since QFTs are expected to provide merely effective descriptions and not to be valid on very small scales (see, e.g., \cite{Wal18} for discussion), this problem is usually neither regarded as unexpected nor as threatening. Still, the problem has attracted much interest over the years, and has been investigated with some success, as one can take a limit of removing the cut-off through a renormalization procedure for some QFTs. We report here that the UV divergence problem does not occur if the Hamiltonian is defined in a novel way that proceeds directly without renormalization.\footnote{When this article was first written (and posted as \url{http://arxiv.org/abs/1505.04847}) in 2015, it was the first to make the proposal to use IBCs to remove ultraviolet divergences. In the meantime, other articles on this topic were published, and we have decided to review some of them in this updated version.}

We describe these Hamiltonians for some non-relativistic QFTs and
report about recent proofs \cite{ibc2a,LS18,Lam18} showing that the Hamiltonians obtained in this way are well-defined and self-adjoint. In this approach, space is continuous (as opposed to a lattice), and the radius of the electron (or other particles) is zero. 
The key element of the approach is a new type of boundary condition 
that we call an \emph{interior--boundary condition} (IBC) because it relates the values of $\psi$ on the boundary of configuration space $\conf$ to the values in the interior of $\Q$, as we will explain presently. As such, IBCs have already been considered in the past, in one form or another, starting with Landau and Peierls \cite{LP30} in 1930, followed by \cite{Mosh51a,Mosh51b, Mosh51c,Tho84,ML91, Yaf92,Tum04} (see below for a brief overview). 
Our new insight is that IBCs, when used for defining QFTs, can avoid the problem of UV divergence. In the past, IBCs received 
little attention, were not explored systematically, were not studied rigorously except to a rather limited extent, and were not considered for addressing UV divergence. 

\begin{figure}[h]
\begin{center}
\includegraphics[width=.7 \textwidth]{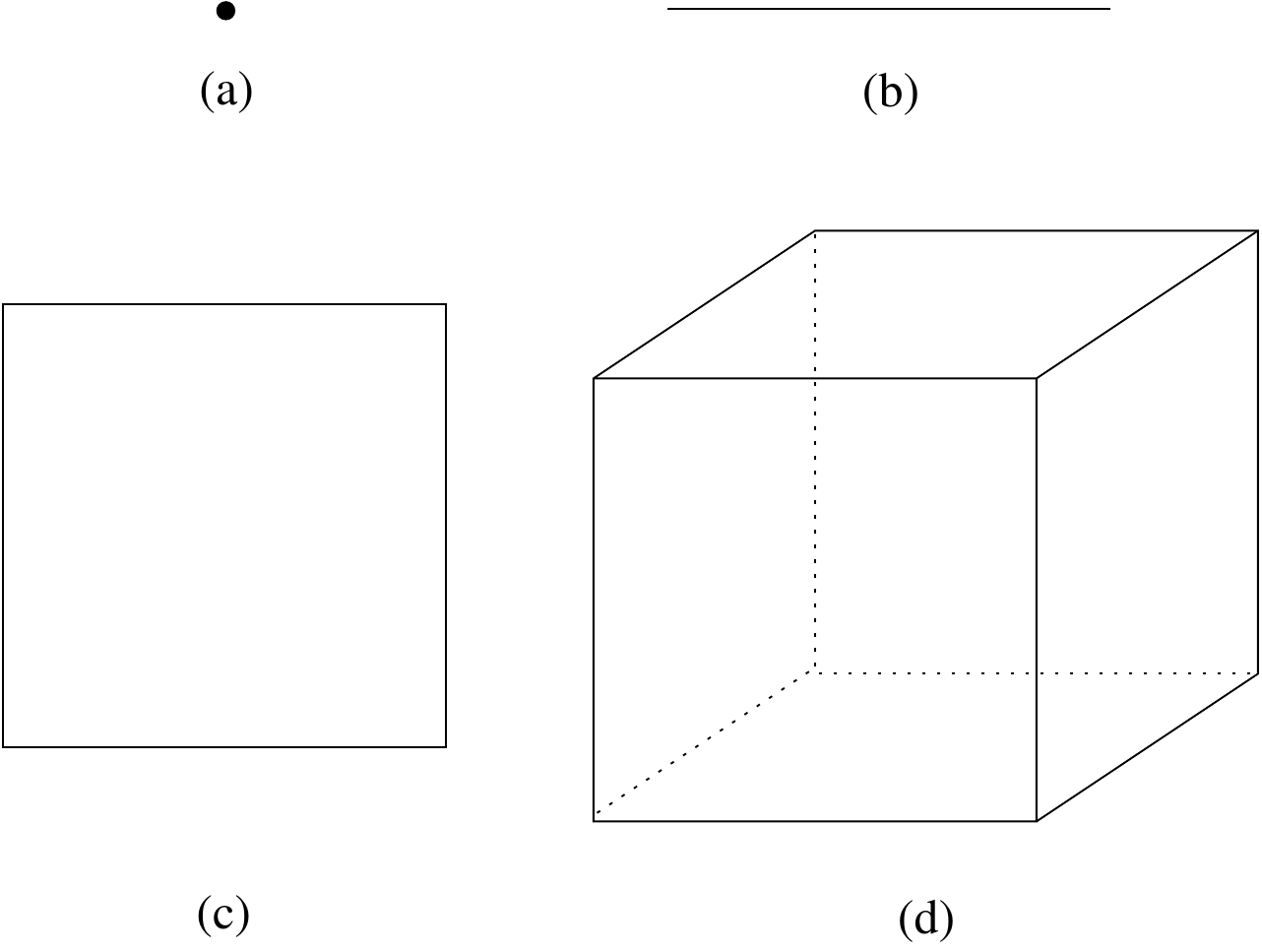}
\end{center}
\caption{Illustration of a configuration space for a variable number of particles in 1 dimension: (a) The zero-particle sector contains only one configuration, the empty configuration; (b) the one-particle sector is a copy of physical space; (c) the two-particle sector; (d) the three-particle sector. In this example, we have not removed the diagonal $\Delta_n$.}
\label{figone}
\end{figure}

IBCs as we use them are formulated in the \emph{particle--position representation} of the state vector $\psi$ in Hilbert space $\Hilbert$. Here, ``particle representation'' means that $\Hilbert$ is represented as a Fock space (or, if appropriate, a tensor product of several Fock spaces), and ``position representation'' that the contribution from the $n$-particle sector of Fock space is represented (like a wave function in quantum mechanics) as a function of $n$ points in 3-dimensional physical space. Specifically, if $\Hilbert$ is a single Fock space then $\psi\in\Hilbert$ can be viewed as a function on a configuration space of a variable number of particles, such as
\be
\Q=\bigcup_{n=0}^\infty \Q_n =\bigcup_{n=0}^\infty\Bigl[(\RRR^3)^n \setminus \Delta_n\Bigr]\,,
\ee
see Figure~\ref{figone}, where
\be
\Delta_n = \Bigl\{ (\vx_1,\ldots,\vx_n)\in (\RRR^3)^n: \vx_i=\vx_j \text{ for some }i\neq j\Bigr\}
\ee
is the ``diagonal,'' i.e., the set of collision configurations (i.e., those with two or more particles at the same location). The relevant boundary $\partial\Q$ of $\Q$ is $\cup_{n=0}^\infty \Delta_n$; the IBC relates the values of $\psi$ on $\partial\Q_n=\Delta_n$ to the values of $\psi$ in the interior of $\Q_{n-1}$, namely at the configuration with one particle removed (or possibly with more than one particle removed, if more than two particles collide). 

The IBC approach allows a nonzero flux of probability out of or into the boundary while a term in the Hamiltonian ensures that all of the $|\psi|^2$ probability lost in one sector gets added in another sector, so that total probability in $\Q$ is conserved. The IBC can thus be regarded as enabling the creation of a contribution to $\psi$ on $\Q_n$ that flows out of the boundary $\partial \Q_n$ and depends on the values of $\psi$ on $\Q_{n-1}$. The approach has been shown in the non-relativistic setting to provide a well-defined Hamiltonian $H_{\mathrm{IBC}}$ whose domain consists of functions satisfying the IBC, without the need for a UV cutoff or renormalization.

Moreover, as we will explain, $H_{\mathrm{IBC}}$ provides a natural mathematical interpretation of the original (UV divergent) expression $H_{\orig}$ for the Hamiltonian. One way of expressing the problem with $H_{\orig}$ is to say that, due to divergent creation terms, $H_{\orig}\psi$ exists as a distribution (such as the Dirac delta distribution) but not as a square-integrabe function, so that $H_{\orig}$ does not define an operator in Hilbert space. The IBC approach actually shares the divergent creation term, while the IBC forces wave functions to diverge at the boundary $\partial \Q_n$ at a particular rate, with the consequence that the Laplacian of $\psi$ also exists as a distribution but not as a square-integrable function. The IBC ensures that the two distributional contributions to $H_{\mathrm{IBC}}\psi$ (one from divergent creation terms and one from the Laplacian acting on a diverging $\psi$) exactly cancel each other, so that their sum is actually a square-integrable function. In fact, the IBC characterizes the domain of functions on which this cancelation occurs. For $\psi$ diverging at the boundary, the annihilation terms in $H_{\orig}$ need a separate definition, which the IBC approach provides as well in a natural manner.

After a condition equivalent to an IBC had been considered as early as 1930 \cite{LP30}, this approach was not followed further; instead, much research took for granted that the Hamiltonian of a QFT is the sum of two self-adjoint operators, the free Hamiltonian and the interaction Hamiltonian. This is not so in the IBC approach, where the free Hamiltonian and the interaction Hamiltonian each map square-integrable functions to distributions and thus are not defined as self-adjoint operators in Hilbert space. It is only their sum that is a self-adjoint operator in Hilbert space. In particular, the full Hamiltonian cannot be regarded as a perturbation of the free Hamiltonian. This feature makes a relevant difference, at least for some UV divergence problems.

For some QFTs, it is possible to take a limit, after introducing a UV cutoff, in which the cutoff is removed, possibly after subtracting an infinite constant from the Hamiltonian. It turns out \cite{ibc2a,LS18} that, at least for some models, this limiting Hamiltonian (called the renormalized Hamiltonian) agrees with $H_{\mathrm{IBC}}$; however, this was not visible from the renormalization procedure before considering the IBC.

In this paper, we describe examples of IBCs and how they help define a Hamiltonian, results about the rigorous existence and self-adjointness of the Hamiltonians, and how these Hamiltonians are related to some known cases in which a UV cut-off can be removed, thus making it plausible that the IBC Hamiltonians are physically relevant and not merely mathematical curiosities. Specifically, we consider two models: In Model 1, $x$-particles can emit and absorb $y$-particles, and both kinds of particles are non-relativistic; we give the full definition in Section~\ref{sec:twomodels} below. The $x$-particles will also be called the ``sources.'' Model 2 is a simplified version of Model 1 in which the $x$-particles cannot move but are fixed at certain locations; it arises as a limiting case of Model 1 in which the mass $m_x$ of the $x$-particle tends to $\infty$; for simplicity, we consider only a single $x$-particle (and call its location the origin). That is, $y$-particles can be created and annihilated at the origin of $\RRR^3$. For Model 2 (the one with fixed $x$-particles), it is known that, after a UV cut-off is introduced, the cut-off can be removed by means of a suitable limiting procedure (renormalization), and we can show that the limiting Hamiltonian $H_\infty$ thus obtained coincides with $H_{\mathrm{IBC}}$ up to addition of a finite constant (see Section~\ref{sec:removecutoff} below). For Model 1, however, such a procedure was not known before the IBC approach succeeded (although a renormalization procedure was known for a similar model, the Nelson model \cite{Nel64}). The IBC approach provides, also for Model 1, a well-defined and self-adjoint Hamiltonian without renormalization \cite{Lam18}.

Let us give a brief overview of prior works considering IBCs.
Historically, the first equation equivalent to an IBC was written down, as far as we know, by Landau and Peierls in 1930 \cite{LP30} when trying to formulate quantum electrodynamics in the particle--position representation, although their Hamiltonian was still UV divergent. Moshinsky~\cite{Mosh51a,Mosh51b} considered, as an effective description of nuclear reactions, a model using the Laplace operator with an IBC for (fixed sources and) a mini-Fock space having only sectors with $n=0$ and $n=1$ particles. Yafaev~\cite{Yaf92} independently considered the same model and proved for that Hamiltonian that it is well defined and self-adjoint. Thomas~\cite{Tho84} considered a similar model with moving sources, also with only two sectors. Georgii and Tumulka~\cite[Sec.~6]{Tum04} considered IBCs for boundaries of codimension~1; such IBCs are developed systematically in \cite{co1}. Recent and upcoming works exploring various aspects of IBCs include \cite{ibc2a,LS18,Lam18,TT15b,KS15,Gal16,LN18,bohmibc,ST18,Sch18,IBCdiracCo1}. 

We give a gentle introduction to IBCs and the ideas behind them in \cite{TT15b}. Mathematical proofs of our main claims are provided in \cite{ibc2a,LS18,Lam18}; the case of fixed sources in 3d is treated in \cite{ibc2a}, the case of moving sources in 2d in \cite{LS18}, and that of moving sources in 3d in \cite{Lam18}. In \cite{LS18}, it is also shown how to generalize the IBC approach to a relativistic dispersion relation $E=\sqrt{m^2c^4+\vp^2c^2}$ (for scalar, not Dirac wave functions) instead of $E=\vp^2/2m$; the condition replacing the IBC is then no longer literally a boundary condition and therefore called an abstract boundary condition. Keppeler and Sieber \cite{KS15} study IBCs in one space dimension. Galvan \cite{Gal16} has proposed an approach similar to IBCs. A goal for the future is to apply the IBC approach to quantum electrodynamics and other serious theories; work on implementing IBCs for the Dirac equation is in progress \cite{LN18,IBCdiracCo1}. For the time being, we report results for non-relativistic model QFTs, based on the Laplace operator.

This paper is organized as follows. In Section~\ref{sec:twomodels}, we give a more detailed description of Model 1 and Model 2, two simple models of non-relativistic QFT. 
In Section~\ref{sec:IBC}, we describe for each of the two models how to set up an IBC and the corresponding Hamiltonian. In Section~\ref{sec:conclusions}, we conclude.

\section{Two Models of Non-Relativistic QFT}
\label{sec:twomodels}

In this section, we describe Model 1 and Model 2, two simple non-relativistic QFTs. 
We set up their Hilbert spaces $\Hilbert$, describe their original Hamiltonians $H_\orig$, and explain why they are UV divergent. Model 1 is a natural, physically reasonable model, while Model 2 is an artificially simplified version that will be useful for an easy discussion of IBCs. In Section~\ref{sec:IBC}, we will describe, for each of the two models, our alternative definition of the Hamiltonian involving an interior--boundary condition.

\subsection{Model 1: $x$-Particles Emit and Absorb $y$-Particles}
\label{sec:ex}
\label{sec:modelAmove}

Model 1 is a QFT related to the Lee model \cite{Lee54}, Schweber's scalar field model \cite[p.~339]{Schw61}, and the Nelson model \cite{Nel64}. It involves two species of particles, $x$ and $y$; the $x$-particles can emit and absorb $y$-particles. Both species are spinless and assume the non-relativistic dispersion relation $E=\vp^2/2m$ with masses $m_x,m_y>0$. The Hilbert space is a tensor product of Fock spaces,
\be\label{Hilbertdef}
\Hilbert = \Fock^- \otimes \Fock^+
\ee
with
\be\label{Fockdef}
\Fock^{\pm} = \bigoplus_{n=0}^\infty S_{\pm} L^2(\RRR^3,\CCC)^{\otimes n}\,,
\ee
where $S_-$ is the anti-symmetrization operator, $S_+$ is the symmetrization operator, and $S_{\pm} L^2(\cdots)$ are their ranges (i.e., the spaces of (anti-)symmetric functions on $(\RRR^3)^n$). Here, we take $x$-particles to be fermions and $y$-particles to be bosons.\footnote{This choice is contrary to the spin--statistics theorem; but that does not matter for our purposes, as the latter pre-supposes Lorentz invariance.} $\RRR^3$ is understood as physical space, i.e., ordinary position space. In the following, we simply write $L^2(\RRR^3)$ for $L^2(\RRR^3,\CCC)$. A vector $\Psi\in\Hilbert$ can be regarded as a function
\be
\psi:\conf_x\times \conf_y \to \CCC
\ee
with $\conf_x=\conf_y$ the configuration space of a variable number of particles,
\be\label{confdef}
\conf_{x}=\conf_y = \bigcup_{n=0}^\infty \conf_n = \bigcup_{n=0}^\infty (\RRR^3)^n\,,
\ee
where the union is understood as a disjoint union and $(\RRR^3)^0=\{\emptyset\}$. We will discuss removing the collision configurations later. 

We call the function $\psi$ the \emph{particle-position representation} of the vector $\Psi\in\Hilbert$. A generic element of $\conf_x\times \conf_y$ can be written as $(x,y)=(\vx_1,\ldots, \vx_m, \vy_1,\ldots,\vy_n)$, where bold-face symbols denote vectors in 3-space, while $x$ denotes a configuration of $x$-particles and $y$ one of $y$-particles; we will often write $x^m$ instead of $x$ to convey that the configuration consists of $m$ $x$-particles, and likewise $y^n$ instead of $y$. We call $(\RRR^3)^m$ the \emph{$m$-particle sector} of $\conf_x$ and $(\RRR^3)^m\times(\RRR^3)^n$ the \emph{$(m,n)$-particle sector} of $\conf_x\times\conf_y$. Likewise, we say \emph{$(m,n)$-particle sector} of $\Hilbert$ (or of $\Psi$, or of $\psi$) and write $\Hilbert^{(m,n)}$ (or $\Psi^{(m,n)}$ or $\psi^{(m,n)}$) for $S_- L^2(\RRR^3,\CCC)^{\otimes m} \otimes S_+ L^2(\RRR^3,\CCC)^{\otimes n}$, respectively for the projection of $\Psi$ to that subspace, and for the restriction of $\psi$ to $(\RRR^3)^m\times(\RRR^3)^n$. The $\psi^{(m,n)}$ function is anti-symmetric in the $x$ variables and symmetric in the $y$ variables.

The spaces $\Q_{x},\Q_y$ are equipped with the volume measure
\be\label{mudef}
\mu(S)= \sum_{n=0}^\infty  \mathrm{vol}_{3n} (S\cap\conf_{n}) \quad \text{for } S\subseteq \Q_x\,.
\ee
The inner product in $\Hilbert$ is then given by
\begin{align}
\scp{\psi}{\phi} &= \int_{\Q_x} \mu(dx) \int_{\Q_y} \mu(dy) \, \psi^*(x,y) \, \phi(x,y)\\
&= \sum_{m=0}^\infty \sum_{n=0}^\infty \Bscp{\psi^{(m,n)}}{\phi^{(m,n)}}_{L^2(\RRR^{3m+3n})}\,.
\end{align}

\subsubsection{Original Hamiltonian}
\label{sec:origH}

As our example of a (non-relativistic) Hamiltonian $H_\orig$, we take
\begin{align}
(H_\orig \psi)^{(m,n)}(x^m,y^n) &= -\frac{\hbar^2}{2m_x} \sum_{i=1}^m \nabla_{\vx_i}^2 \psi^{(m,n)} (x^m,y^n)
-\frac{\hbar^2}{2m_y} \sum_{j=1}^n \nabla_{\vy_j}^2 \psi^{(m,n)}(x^m,y^n) \nonumber\\
&\quad +\: nE_0 \psi^{(m,n)}(x^m,y^n) \nonumber\\
&\quad +\: g \sqrt{n+1} \sum_{i=1}^m \psi^{(m,n+1)}\bigl(x^m, (y^n,\vx_i)\bigr)\nonumber\\
&\quad +\: \frac{g}{\sqrt{n}} \sum_{i=1}^m \sum_{j=1}^n  \delta^3(\vx_i-\vy_j)\,\psi^{(m,n-1)}\bigl(x^m,y^n\setminus \vy_j\bigr)\,,\label{Horigdef1}
\end{align}
using the notation $\delta^3$ for the 3-dimensional Dirac delta function and
\be\label{ysetminusdef}
y^n\setminus \vy_j = (\vy_1,\ldots,\vy_{j-1},\vy_{j+1},\ldots,\vy_n)
\ee
for the configuration of $n-1$ $y$-particles with the $j$-th particle removed; $E_0$ is the energy that must be expended for creating a $y$-particle (the ``$y$ rest energy''), and $g\in\RRR$ is a coupling constant (i.e., the ``charge'' of an $x$-particle).
Instead of $\psi^{(m,n)}(x^m,y^n)$, we can also simply write $\psi(x^m,y^n)$, as the argument uniquely determines which sector of $\psi$ must be used; we sometimes find it useful to use the more explicit notation to make the relations to other sectors more easily visible.
The first term in \eqref{Horigdef1} is the free fermion Hamiltonian $H_x\psi$, the second and third terms are the free boson Hamiltonian $H_y\psi$, and the third and fourth line together are the interaction Hamiltonian $H_\inter\psi$ responsible for the creation and annihilation of $y$-particles. In terms of creation and annihilation operators,
\begin{align}
H_{x} &= \frac{\hbar^2}{2m_{x}}\int d^3\vq \, \nabla a_{x}^\dagger(\vq) \, \nabla a_{x}(\vq)\\
H_{y} &= \frac{\hbar^2}{2m_{y}}\int d^3\vq \, \nabla a_{y}^\dagger(\vq) \, \nabla a_{y}(\vq) 
+ E_0 \int d^3\vq\, a_y^\dagger(\vq)\, a_y(\vq)\\
H_\inter &= g \int d^3\vq\, a_x^\dagger(\vq) \, \bigl(a_y(\vq)+a^\dagger_y(\vq) \bigr)\, a_x(\vq)
\end{align}
with $^\dagger$ denoting the adjoint operator, and $a_{x}(\vq)$, $a_{y}(\vq)$ the annihilation operators for an $x,y$-particle at location $\vq$ in position space, formally defined by
\begin{align}
\bigl(a_x(\vq)\,\psi\bigr)(x^m,y^n) &= \sqrt{m+1}\, \psi^{(m+1,n)}\bigl((x^m,\vq),y^n\bigr)\label{axdef}\\
\bigl(a_x^\dagger(\vq)\,\psi\bigr)(x^m,y^n) &= \frac{1}{\sqrt{m}} \sum_{i=1}^m (-1)^i \,\delta^3(\vx_i-\vq)\, \psi^{(m-1,n)}\bigl(x^m\setminus \vx_i,y^n\bigr)\,,\\
\bigl(a_y(\vq)\,\psi\bigr)(x^m,y^n) &= \sqrt{n+1}\, \psi^{(m,n+1)}\bigl(x^m,(y^n,\vq)\bigr)\label{aydef}\\
\bigl(a_y^\dagger(\vq)\,\psi\bigr)(x^m,y^n) &= \frac{1}{\sqrt{n}} \sum_{j=1}^n  \delta^3(\vy_j-\vq)\,\psi^{(m,n-1)}\bigl(x^m,y^n\setminus \vy_j\bigr)\,.
\end{align}
The combinatorial factors $\sqrt{m+1}$ and $m^{-1/2}$ arise from the fact that every configuration of $m$ $x$-particles occurs in $m!$ different permutations.  Since the operators $a_{x,y}(\vq)$ contain the point evaluation of functions, they can actually be defined  rigorously on a dense subspace of $\Hilbert$, while their adjoints are ill defined for any nonzero $\psi$.

Model 1 is not Galilean covariant but a modification that is can be set up \cite{LL67} and can be combined with IBCs \cite[Section~4.2]{bohmibc}.

\subsubsection{UV Divergence of the Original Hamiltonian}

The original Hamiltonian $H_\orig$ as defined in \eqref{Horigdef1} above is UV divergent and thus ill defined. The source of the difficulty is the delta function in the last line of \eqref{Horigdef1}: Since a delta function is not a square-integrable function, the right-hand side of \eqref{Horigdef1} does not lie in $\Hilbert$ for any choice of $\psi$; thus, \eqref{Horigdef1} does not define an operator in $\Hilbert$. It can also be pointed out that the delta function $\delta^3$ (which plays the role in \eqref{Horigdef1} of the wave function of a newly created $y$-particle) is a state of infinite energy, 
\be
\text{formally} \quad
\Bigl\langle \delta^3 \Big| \Bigl( -\frac{\hbar^2}{2m_y} \nabla^2 \Bigr) \Big| \delta^3 \Bigr\rangle =\int_{\RRR^3} d^3\vk \, \frac{(\hbar\vk)^2}{2m_y} =\infty\,.
\ee

The standard procedure for obtaining a well-defined Hamiltonian (UV cut-off) is to replace the delta function by a square-integrable function $\varphi:\RRR^3\to\CCC$, yielding
\begin{align}
(H_\cutoff \psi)(x^m,y^n) &= -\frac{\hbar^2}{2m_x} \sum_{i=1}^m \nabla_{\vx_i}^2 \psi (x^m,y^n)
-\frac{\hbar^2}{2m_y} \sum_{j=1}^n \nabla_{\vy_j}^2 \psi(x^m,y^n) \nonumber\\
&\quad +\: nE_0 \psi(x^m,y^n) \nonumber\\
&\quad +\: g \sqrt{n+1} \sum_{i=1}^m \int_{\RRR^3} d^3\vy\, \varphi^*(\vx_i-\vy)\, \psi\bigl(x^m, (y^n,\vy)\bigr) \nonumber\\
&\quad +\: \frac{g}{\sqrt{n}} \sum_{i=1}^m \sum_{j=1}^n \varphi(\vx_i-\vy_j) \,\psi\bigl(x^m,y^n\setminus \vy_j\bigr) \,,\label{Hcutoffdef}
\end{align}
which amounts to saying that an $x$-particle has an extended charge distribution with density function $\varphi$ (if $\varphi$ is real-valued). Equivalently, the interaction Hamiltonian is replaced by
\be
H_{\inter,\cutoff} = g \int d^3\vq\, a_x^\dagger(\vq) \, \bigl(a_{y,\varphi}(\vq)+a^\dagger_{y,\varphi}(\vq) \bigr)\, a_x(\vq)\,,
\ee
where
\begin{align}
\bigl(a_{y,\varphi}(\vq)\, \psi\bigr)(x^m,y^n) &= \sqrt{n+1}\int_{\RRR^3} d^3\vy\, \varphi^*(\vq-\vy)\, \psi^{(m,n+1)}\bigl(x^m,(y^n,\vy)\bigr)\\
\bigl(a_{y,\varphi}^\dagger(\vq)\,\psi\bigr)(x^m,y^n) &= \frac{1}{\sqrt{n}} \sum_{j=1}^n  \varphi(\vy_j-\vq)\,\psi^{(m,n-1)}\bigl(x^m,y^n\setminus \vy_j\bigr)\,.
\end{align}

\subsection{Simplified Version: Model 2}
\label{sec:modelAfixed}

In Model 2, there is only one $x$-particle, and it is fixed at the origin. The only dynamical (and quantized) degrees of freedom reside in the $y$-particles; such models were considered in particular by van Hove \cite{vH52,Der03} and Lee \cite{Lee54}, and they tend to arise in the limit $m_x\to\infty$. Specifically, we take the configuration space of Model 2 to be the set 
\be
\Q=\Q_y=\bigcup_{n=0}^\infty \Q^{(n)}_y= \bigcup_{n=0}^\infty (\RRR^3\setminus \{\vzero\})^n\,,
\ee
where $\vzero$ denotes the origin in $\RRR^3$; a configuration is denoted by $y^n=(\vy_1,\ldots,\vy_n)$. Note that we exclude the possibility that any $y$-particle can be at the location of the $x$-particle (i.e., the origin), but we do not exclude the possibility that two $y$-particles can be at the same location because for the purposes of this model, in which the $y$-particles do not interact, there is no need to exclude such configurations.

Correspondingly, we take the Hilbert space to be the bosonic Fock space
\be
\Hilbert =\Fock^+= \bigoplus_{n=0}^\infty S_+ L^2(\RRR^3)^{\otimes n}\,.
\ee
Since for the definition of the Hilbert space it plays no role whether the origin is excluded or not, elements $\psi\in\Hilbert$ can be regarded as complex-valued functions on $\Q$ that are permutation-symmetric in every sector; thus, $\Hilbert \subset L^2(\Q)$, where $\Q$ is thought of as equipped with the measure $\mu$ defined in the same way as in \eqref{mudef} ($L^2(\Q)$ contains also non-symmetric functions). The ``original'' Hamiltonian is now a simplified version of \eqref{Horigdef1}:
\begin{align}
(H_\orig \psi)^{(n)}(y^n) &= -\frac{\hbar^2}{2m_y} \sum_{j=1}^n \nabla_{\vy_j}^2 \psi^{(n)}(y^n) + nE_0 \psi^{(n)}(y^n) \nonumber\\
&\quad +\: g \sqrt{n+1} \, \psi^{(n+1)}(y^n,\vzero)\nonumber\\
&\quad +\: \frac{g}{\sqrt{n}} \sum_{j=1}^n  \delta^3(\vy_j)\,\psi^{(n-1)}\bigl(y^n\setminus \vy_j\bigr)\,,\label{Horigdef9}
\end{align}
Also this Hamiltonian is UV divergent.

\section{Interior--Boundary Condition and Corresponding Hamiltonian for the Two Models}
\label{sec:IBC}

For Model 1 and Model 2, we describe the IBC and the Hamiltonian $H_{\mathrm{IBC}}$. We also explain why $H_{\mathrm{IBC}}$ is a reasonable interpretation of the formula for $H_\orig$. We begin with the simpler scenario of Model 2.

\subsection{IBC for Model 2}

Let $\SSS^2$ denote the unit sphere in $\RRR^3$. 
The IBC demands the following: For every $\vomega\in\SSS^2$, $n\in\{0,1,2,\ldots\}$, $y^n\in(\RRR^3\setminus\{\vzero\})^n$,
\be\label{IBC9a}
\lim_{r\searrow 0} \biggl(r\psi^{(n+1)}(y^n,r\vomega)  \biggr) 
= - \frac{g\, m_y}{2\pi\hbar^2\sqrt{n+1}}\psi^{(n)}(y^n)\,.
\ee
The IBC is a condition on the wave function $\psi$ at or near the ``boundary'' of configuration space; the relevant boundary $\partial\Q_y^{(n)}$ of $\Q_y^{(n)}$ is 
\be
\partial \Q_y^{(n)} = \bigcup_{j=1}^n (\RRR^3)^{j-1} \times \{\vzero\} \times (\RRR^3)^{n-j}\,.
\ee
That is, the boundary consists of those configurations at which one of the $y$-particles collides with the $x$-particle. Due to the permutation symmetry of $\psi^{(n+1)}$, we can assume without loss of generality that it is the $n+1$-st variable, $\vy_{n+1}$, that approaches $\vzero$. The name ``interior--boundary condition'' reflects the fact that \eqref{IBC9a} relates the values (or limits) of $\psi$ on the boundary $\partial \Q_y$ to values of $\psi$ in the interior of configuration space $\conf$ (namely, in the interior of a different sector corresponding to a lesser number of $y$-particles).

The IBC \eqref{IBC9a} allows in particular that $\psi(y^n,r\vomega)$ diverges like $1/r$ as $r\to 0$. In fact, it requires that $\psi$ so diverges whenever the right-hand side of \eqref{IBC9a} is nonzero.\footnote{\label{fn:1/r}This divergence is to be expected if we keep in mind that $|\psi|^2$ represents probability density and note that for the inward radial flow in $\RRR^3$, i.e., for the dynamical system defined by the ODE $\dot{\vx} = -\vx/|\vx|$, the stationary (rotationally invariant) density is $1/r^2=1/|\vx|^2$. To see this, note that if we squeeze a spherical shell of radius $r_1$ and thickness $dr$ to a shell with smaller radius $r_2$ and equal thickness $dr$ then its volume goes down (and thus, if its probability content is conserved, its density goes up) by a factor of $(r_1/r_2)^2$. That is a reason for expecting $|\psi|^2\sim 1/r^2$ and thus $\psi\sim 1/r$ as $r\to 0$. We note, however, that this reasoning depends on the radial velocity being constant (or becoming constant as $r\to 0$); this turns out true in 3 space dimensions but not in 2 \cite{LS18}, so the the reasoning suggests the wrong asymptotics in 2d; see \cite[Rem.~13]{bohmibc}.} By permutation symmetry, $\psi(y^n)$ diverges as \emph{any} $\vy_j$ approaches $\vzero$.

On wave functions $\psi$ satisfying the IBC \eqref{IBC9a}, the Hamiltonian $H=H_{\mathrm{IBC}}$ is defined by
\begin{align}
(H_{\mathrm{IBC}}\psi)^{(n)}(y^n) 
&= -\frac{\hbar^2}{2m_y} \sum_{j=1}^{n} \nabla^2_{\vy_j}\psi^{(n)}(y^n)+ nE_0 \psi^{(n)}(y^n) \nonumber\\[2mm]
&\quad +\: \frac{g\sqrt{n+1}}{4\pi}\int\limits_{\SSS^2} d^2\vomega \, \lim_{r\searrow 0} \frac{\partial}{\partial r} \Bigl( r \psi^{(n+1)}(y^n,r\vomega) \Bigr)\nonumber\\
&\quad +\: \frac{g}{\sqrt{n}} \sum_{j=1}^n  \delta^3(\vy_j)\,\psi^{(n-1)}(y^n\setminus \vy_j)\,.\label{Hdef9a}
\end{align}
This equation differs from the expression \eqref{Horigdef9} for $H_{\orig}$ only in the second line, where $\psi(y^n,\vzero)$ has been replaced by a more complicated expression involving the behavior of $\psi$ near the configuration $(y^n,\vzero)$; after all, $\psi$ diverges at this configuration by virtue of the IBC, so the expression $\psi(y^n,\vzero)$ does not make sense. The great similarity between $H_{\mathrm{IBC}}$ and $H_{\orig}$ adds to the suggestion that $H_{\mathrm{IBC}}$ is the right Hamiltonian because it is a mathematical interpretation of the formal expression $H_{\orig}$.

In view of the delta function appearing in \eqref{Hdef9a}, it may seem unlikely that such a Hamiltonian can be well defined. However, a theorem from \cite{ibc2a}, repeated here as Theorem~\ref{thm:fixedx} in Section~\ref{sec:rigorous} below, shows that in fact it is well defined. It turns out that 
the last line in \eqref{Hdef9a} always gets canceled by contributions to the first line, which may contain delta functions because we now allow wave functions that diverge like $1/r=1/|\vy_j|$, and the Laplacian of $1/r$ is $-4\pi\delta^3$ (as readers may recall from electrostatics, where $1/r$ occurs as the Coulomb potential $\phi$ generated by a point charge, satisfying the Poisson equation $\Delta \phi = -4\pi \rho$ with charge density $\rho$). As a consequence of this cancelation, $H_{\mathrm{IBC}}\psi$ is a square-integrable function for $\psi$ satisfying the IBC. Conversely, if we want the Laplacian to produce a contribution that is a Dirac delta contribution, we are led to allowing $\psi$ that diverges like $1/r$, and then the necessary and sufficient condition for the cancelation is just the IBC. See also Remark~\ref{rem:Diracdelta} in Section~\ref{sec:rem} below for further discussion.

Let us point out a connection between the second line of \eqref{Hdef9a}, the line that differed from the original Hamiltonian \eqref{Horigdef9}, and the corresponding line in \eqref{Horigdef9}. Think of $\psi\bigl(y^n, r\vomega\bigr)$ as a function of $r$; as $r\to 0$, a $\psi$ from the domain of $H_{\mathrm{IBC}}$ can be expanded in the form
\be\label{model2asymp}
\psi(y^n,r\vomega) = \frac{c_{-1}(y^n)}{r} + c_0(y^n) + O(r)
\ee
with complex coefficients $c_{-1}(y^n),c_0(y^n)$; moreover,
\be\label{dpowersofr}
\frac{\partial}{\partial r}\bigl(r\psi(y^n,r\vomega) \bigr) = c_0(y^n)+O(r)\,, 
\ee
which yields $c_0$ in the limit $r\to 0$. If $\psi$ did not diverge as $r\to 0$, this would be exactly the value $\lim_{r\searrow 0} \psi(y^n,r\vomega) = \psi\bigl(y^n,\vzero\bigr)$ occurring in \eqref{Horigdef9} in the second line. So, \eqref{Hdef9a} is really quite similar to \eqref{Horigdef9}. Indeed, we can regard $c_0$ as the natural interpretation of the expression $\psi(y^n,\vzero)$ in a situation in which $\psi$ also contains a term $c_{-1}/r$ diverging at $\vzero$; put differently, it is a way of extending the annihilation operator $a_y(\vzero)$ to functions containing a term $c_{-1}/r$. Thus, in total, $H_{\mathrm{IBC}}$ appears as a natural interpretation of $H_{\orig}$, with the IBC characterizing the right domain on which the distribution-valued terms in $H_{\mathrm{IBC}}\psi$ cancel each other, leaving a square-integrable function.

Keppeler and Sieber \cite{KS15}, apart from developing a 1-dimensional version, give further reasons why the IBC model represents a reasonable interpretation of the original Hamiltonian \eqref{Horigdef9}. They argue \cite[App.~A]{KS15} that for any eigenfunction $\psi$ of \eqref{Horigdef9}, when integrating \eqref{Horigdef9} in $\vy_{n}$ over a ball of radius $r$ around the source, using the Gauss integral theorem on the Laplacian term, and taking the limit $r\to 0$, one should obtain that 
\be
0=-\frac{\hbar^2}{2m_y} \lim_{r\searrow 0} \int_{\SSS^2} d^2\vomega\, r^2\, \frac{\partial}{\partial r}\psi^{(n)}(y^{n-1},r\vomega)
+ \frac{g}{\sqrt{n}} \psi^{(n-1)}(y^{n-1})\,.
\ee
Assuming that $\psi^{(n)}$ diverges as $r\to0$ no worse than $1/r$, this relation is equivalent to the IBC \eqref{IBC9a}. Finally, if every eigenfunction satisfies the IBC \eqref{IBC9a}, then so does any function in the domain of the Hamiltonian. The corresponding reasoning in the 1-dimensional case yields that $\psi$ should have a jump in the first derivative at the origin, with the magnitude of the jump proportional to the wave function in the next lower sector.

The Hamiltonian $H_{\mathrm{IBC}}$ is \emph{not} the sum of two self-adjoint operators, the free Hamilonian and an interaction Hamiltonian. That is because the free Hamiltonian is defined on a different domain than $H_{\mathrm{IBC}}$, containing wave functions that do not satisfy the IBC and do not diverge on the diagonal. It is the action of the Laplacian on functions that diverge on the diagonal that leads to delta functions and thus makes the last line of \eqref{Hdef9a} possible, and it is only in conjunction with the IBC that the Hamiltonian \eqref{Hdef9a} leads to conservation of probability. 

Indeed, in order to understand why the IBC \eqref{IBC9a} was chosen this way, and how it works together with the formula \eqref{Hdef9a} for the Hamiltonian, it is illuminating to calculate the balance equation for $|\psi|^2$ and check that $|\psi|^2$ is conserved. We go through such a calculation in the next section.

\subsection{Self-Adjointness and Conservation of Probability}
\label{sec:conservation}
\label{sec:rigorous}

\begin{thm}\label{thm:fixedx} {\rm \cite{ibc2a}}
On a certain dense subspace $\domain_{\mathrm{IBC}}$ of $\Hilbert=\Fock^+$, the elements of which satisfy the IBC \eqref{IBC9a}, the operator $H_{\mathrm{IBC}}$ given by \eqref{Hdef9a} 
is well-defined and self-adjoint. If $E_0\geq 0$, then $H_{\mathrm{IBC}}$ is bounded from below.
\end{thm}

The relevance of this theorem is that it makes clear that $H_{\mathrm{IBC}}$ is not afflicted by UV divergence. We give here a calculation checking on a non-rigorous level that probability is conserved. Using the symbol
\be
\vj_{\vy_j}=\vj_{\vy_j}(y^n)= \frac{\hbar}{m_y} \Im \, \psi^* \nabla_{\vy_j} \psi
\ee
for the usual probability current, we obtain from \eqref{Hdef9a} that, at any configuration $y^n$ (without any $y$-particle at the origin),
\begin{align}
\frac{\partial \bigl|\psi(y^n)\bigr|^2}{\partial t} 
&= -\sum_{j=1}^n \nabla_{\vy_j}\cdot \vj_{\vy_j} \nonumber\\
&\quad +\: \frac{g\sqrt{n+1}}{2\pi\hbar} \Im \, \psi^*(y^n)\int\limits_{\SSS^2} d^2\vomega \, \lim_{r\searrow 0} \frac{\partial}{\partial r} \Bigl( r \psi(y^n,r\vomega)  \Bigr)\,.
\end{align}
If $g\neq 0$, then the last line can be re-written, using the IBC \eqref{IBC9a}, as
\begin{align}
& -\frac{g\sqrt{n+1}}{2\pi\hbar} \Im \, \frac{2\pi\hbar^2\sqrt{n+1}}{g\,m_y} \lim_{r\searrow 0} \, r\psi^*(y^n,r\vomega)
\int\limits_{\SSS^2} d^2\vomega \, \lim_{r\searrow 0} \frac{\partial}{\partial r} \Bigl( r \psi(y^n,r\vomega)  \Bigr)\\
&= -(n+1)   \lim_{r\searrow 0} \,r^2
\int\limits_{\SSS^2} d^2\vomega \,  \frac{\hbar}{m_y}\Im \, \psi^*(y^n,r\vomega)
 \frac{\partial}{\partial r} \psi(y^n,r\vomega) \,,
\end{align}
using that $\partial_r(r\psi)=\psi + r\partial_r \psi$ and $\Im (r|\psi|^2)=0$. Thus, 
\be
\frac{\partial \bigl|\psi(y^n)\bigr|^2}{\partial t} 
= -\sum_{j=1}^n \nabla_{\vy_j}\cdot \vj_{\vy_j} 
- (n+1)  \lim_{r\searrow 0} \,r^2
\int\limits_{\SSS^2} d^2\vomega \,  \vomega\cdot \vj_{\vy_{n+1}}(y^n,r\vomega) \,.\label{balance1a}
\ee
This equation possesses a simple interpretation: $-r^2$ times the integral is just the flux of probability current toward the origin across the sphere of radius $r$ in the coordinate space of $\vy_{n+1}$; the limit of that as $r\to 0$ is the current of probability into the origin in the coordinate space of $\vy_{n+1}$; summing over all $y$-particles would yield, due to the bosonic symmetry of $\psi$, $n+1$ equal terms; thus, the second summand on the right-hand side of \eqref{balance1a} is the total flux of probability into the boundary configurations obtained from $y^n$ by adding one $y$-particle at the origin. That is, the balance equation \eqref{balance1a} asserts that the probability density $|\psi|^2$ changes in two ways, due to transport of probability in the $n$-particle sector and by increasing at just the rate at which probability disappears on the $(n+1)$-particle sector by flowing into the boundary $\partial \Q_y^{(n+1)}$. Therefore, total probability is conserved.

If $g$ is set to 0, then the amount of probability exchanged between different sectors vanishes, and the norm of each sector, $\int_{\RRR^{3n}} dy^n\, |\psi(y^n)|^2$, is conserved separately, corresponding to the fact that the Hamiltonian $H_{\mathrm{IBC}}$ of \eqref{Hdef9a} commutes with the $y$-particle number operator in this case. In fact, $H_{\mathrm{IBC}}$ reduces to the free Hamiltonian (i.e., the ``second quantization'' of the free 1-particle operator $-(\hbar^2/2m_y)\nabla_{\vy}^2 +E_0$), and the IBC \eqref{IBC9a} asserts that, in the expansion $\psi(y^n,r\vomega) = c_{-1}/r + c_0 + O(r)$ considered in \eqref{dpowersofr}, $c_{-1}=0$; thus, for $g=0$ the IBC demands no more than that $\psi$ is non-singular at $r=0$, as functions in the domain of the free Hamiltonian must be.

\subsection{Remarks}
\label{sec:rem}

\begin{enumerate}

\item \emph{Comparison to Bethe--Peierls boundary condition.} The IBC \eqref{IBC1b} 
has some parallels to the Bethe--Peierls boundary condition \cite{BP35}, which reads for a wave function $\psi\in L^2(\RRR^3,\CCC)$: 
\be\label{BethePeierls}
\lim_{r\searrow 0}\Bigl(\alpha +\frac{\partial}{\partial r} \Bigr) \Bigl(r\psi(r\vomega)\Bigr) =0
\ee
with given constant $\alpha\in\RRR$. This condition is used for giving precise meaning to a Schr\"odinger equation for $\psi:\RRR^3\to\CCC$ with a Dirac delta function as the potential,
\be
H = -\tfrac{\hbar^2}{2m} \Laplace + g\, \delta^3(\vx)\,,
\ee
a kind of interaction known as a \emph{point interaction} \cite{AGHKH88}. Like our IBCs \eqref{IBC9a}, \eqref{IBC9b}, \eqref{IBC1b}, the Bethe--Peierls boundary condition \eqref{BethePeierls} concerns the ``boundary'' at $r=0$. However, in contrast to the IBC, which connects two sectors of $\Q$, the Bethe--Peierls boundary condition involves only one sector, as the wave function $\psi$ is defined on $\RRR^3$. Also, the Bethe--Peierls boundary condition implies zero current into $r=0$ (see, e.g., \cite{TT15b} for more detail), whereas the IBC leads to a nonzero current into the boundary.

\item \emph{Comparison to vertex conditions in networks.}
An example of a known boundary condition leading to nonzero current into the boundary is provided by the vertex conditions for quantum mechanics on networks (see, e.g., \cite{kuchment} and references therein). A network, or graph, consists of several 1-dimensional spaces (i.e., intervals) called edges, glued together at their end points called vertices. The wave function is a (say, complex-valued) function on the network (i.e., on the union of the edges), and the Hamiltonian is given by the Laplace operator on each edge, along with boundary conditions for the end points of the edges, also called vertex conditions. The vertex conditions are related to the fact that, since no positive amount of probability can be located at a vertex, all of the probability current into the vertex must be compensated by a current out of the vertex; that is, for each vertex $v$, the sum of the currents along all edges $e$ connected to $v$ (pointing away from $v$) must vanish,
\be\label{Kirchhoff}
\sum_e j_{e}(v)=0\,.
\ee
While Dirichlet or Neumann boundary condition at every end point of every edge would lead to $j_e(v)=0$ for all $v$ and $e$, it is also possible to impose vertex conditions on $\psi$ that allow nonzero flow of probability from one edge across a vertex to another edge, the simplest one being the conjunction of
\be\label{vertexcont}
\lim_{x\to v \text{ along }e} \psi(x) = \lim_{x\to v \text{ along }e'} \psi(x)\,,
\ee
i.e., that $\psi$ is continuous at vertices, and
\be\label{vertexcond}
\sum_e \partial_e \psi(v) = \alpha \psi(v)\,,
\ee
where $\alpha\in\RRR$ is a given constant and $\partial_e\psi(v)$ means the derivative of $\psi$ along the edge $e$, taken at $v$. 

To draw parallels between \eqref{vertexcond} and IBCs, we may compare the edges of a network to the sectors of the configuration spaces considered in this paper; let us call both the ``blocks'' of the space. Of course, the edges of a graph have equal dimension while the sectors of our $\Q$ have different dimension. We may note that in both situations the probability loss in one block is compensated by a probability gain in others. Also, the condition of continuity \eqref{vertexcont} is a relation between the boundary value in one block and a value in another, but in contrast to an IBC the latter is also a boundary value; that is, in analogy to the terminology ``interior--boundary condition,'' \eqref{vertexcont} is a ``boundary--boundary condition.''

\item \emph{Combinatorial factors.} As remarked at the end of Section~\ref{sec:origH}, the combinatorial factors $\sqrt{n}$ and $\sqrt{n+1}$ that appear in the Hamiltonian arise from the fact that we use \emph{ordered} configurations $(\vy_1,\ldots,\vy_n)$ although in nature configurations are \emph{unordered}, as in $\{\vy_1,\ldots,\vy_n\}$. This is also true of the IBC \eqref{IBC9a} and the associated Hamiltonian \eqref{Hdef9a}. If we used unordered configurations, which leads to topologically non-trivial configuration spaces (see, e.g., \cite{LM77,fermionic}), these factors would not appear.

\item \emph{Neumann-type IBC.} \label{rem:Neumann} Two well known types of boundary conditions are Dirichlet boundary conditions,
\be
\psi\Big|_{\partial \Q}=0\,,
\ee
and Neumann boundary conditions,
\be
\frac{\partial\psi}{\partial n}\Big|_{\partial \Q} = 0\,,
\ee
with $n$ the normal vector on the boundary hypersurface. While the IBC \eqref{IBC9a} resembles the Dirichlet type in that it involves the \emph{values} of $\psi$ (or rather, $r\psi$) on the boundary, one can as well set up a different IBC that resembles the Neumann type in that it involves the \emph{normal derivative} of $\psi$ (or rather, $r\psi$). Here, $r=0$ plays the role of the boundary, and the radial direction plays the role of the normal to the boundary. The \emph{Neumann-type IBC} asserts that for any $n\in\{0,1,2,\ldots\}$, any configuration $y^n\in(\RRR^{3}\setminus\{\vzero\})^n$, and any $\vomega\in\SSS^2$,
\be\label{IBC9b}
\lim_{r\searrow 0} \frac{\partial}{\partial r} \Bigl(r\psi^{(n+1)}(y^n,r\vomega)\Bigr)
= \frac{g\,m_y}{2\pi\hbar^2\sqrt{n+1}} \, \psi^{(n)}(y^n)\,.
\ee
Also this IBC typically leads to $\psi^{(n+1)}(y^n,r\vomega)$ diverging like $1/r$ as $r\to 0$. The name ``Neumann-type IBC'' should be understood with care since, as we have observed already in \eqref{dpowersofr}, the expression $\lim_{r\to 0}\partial_r (r\psi)$ yields, for $\psi=c_{-1}/r+c_0+O(r)$, just $c_0$; as a consequence, in the case $c_{-1}=0$ in which $\psi$ can actually be evaluated at $r=0$, the expression $\lim_{r\to 0}\partial_r (r\psi)$ just yields $\psi(r=0)$, so that the IBC \eqref{IBC9b} has quite the character of a Dirichlet boundary condition.

On wave functions $\psi$ satisfying the IBC \eqref{IBC9b}, the Hamiltonian $H=H_{\mathrm{IBC}}$ is defined by
\begin{align}
(H\psi)^{(n)}(y^n) 
&= -\frac{\hbar^2}{2m_y} \sum_{j=1}^{n} \nabla^2_{\vy_j}\psi^{(n)}(y^n) + nE_0 \psi^{(n)}(y^n) \nonumber\\[2mm]
&\quad + \frac{g\sqrt{n+1}}{4\pi}\int\limits_{\SSS^2} d^2\vomega \, \lim_{r\searrow 0}   r \psi^{(n+1)}(y^n,r\vomega) 
\nonumber\\
&\quad +\: \frac{\hbar^2}{2m_y} \sum_{j=1}^n  \delta^3(\vy_j)
\int\limits_{\SSS^2} d^2\vomega \, \lim_{r\searrow 0} r\psi^{(n-1)}\bigl(y^n\setminus \vy_j,r\vomega\bigr)\,.
\label{Hdef9b}
\end{align}
This equation differs from \eqref{Hdef9a} (for the Dirichlet-type case) in the second line, which no longer involves a derivative, and in the last line, which we will discuss further in 
Remark~\ref{rem:Diracdelta} below.

While \eqref{IBC9b} and \eqref{Hdef9b} together define a different time evolution for $\psi$ than \eqref{IBC9a} and \eqref{Hdef9a}, Equation \eqref{balance1a} is still true and guarantees that the amount of probability lost on the $n+1$-sector of $\Q$ due to probability flux into the boundary is added on the $n$-sector. In fact, as we show in \cite{ibc2a}, also \eqref{IBC9b} and \eqref{Hdef9b} define a self-adjoint operator, provided $E_0>0$.

For $g=0$, it is again true (as in the Dirichlet-type case) that the sectors decouple, and that the $y$-particle number operator is conserved. However, the Hamiltonian does not become the free Hamiltonian. Rather, the IBC \eqref{IBC9b} becomes a Bethe--Peierls boundary condition \eqref{BethePeierls} with $\alpha=0$, and the Hamiltonian becomes the second quantization of the negative Laplacian with a point interaction at the origin.

\item \emph{Robin-type IBC.} \label{rem:Robin} A Robin boundary condition is one of the form
\be
\Bigl(\alpha\psi+ \beta \frac{\partial\psi}{\partial n}\Bigr)\Big|_{\partial \Q} = 0
\ee
with given constants $\alpha$ and $\beta$. An IBC of an analogous form can be set up as follows:
for any $n\in\{0,1,2,\ldots\}$, any configuration $y^n=(\vy_1,\ldots,\vy_{n})\in \bigl(\RRR^{3}\setminus\{\vzero\}\bigr)^n$ of $y$-particles, and any $\vomega\in\SSS^2$,
\be\label{IBC1b}
\lim_{r\searrow 0} \biggl(\alpha+\beta \frac{\partial}{\partial r}\biggr) \biggl(r\psi^{(n+1)}(y^n,r\vomega)\biggr)
= \frac{2m_y}{\hbar^2\sqrt{n+1}} \, \psi^{(n)}(y^n)\,,
\ee
where $(\alpha,\beta)\in \CCC^2\setminus\{(0,0)\}$ are constants. Also this condition typically leads to $\psi$ that diverge like $1/r$ as $r\to0$. The Dirichlet-type condition \eqref{IBC9a} is included in this scheme for $\alpha=-4\pi/g,\beta=0$, while the Neumann-type condition \eqref{IBC9b} is included for $\alpha=0,\beta=4\pi/g$.

The associated Hamiltonian $H=H_{\mathrm{IBC}}$ is given by 
\begin{align}
(H\psi)^{(n)}(y^n) 
&= 
-\frac{\hbar^2}{2m_y} \sum_{j=1}^{n} \nabla^2_{\vy_j}\psi^{(n)}(y^n) + nE_0 \psi^{(n)}(y^n) \nonumber\\[2mm]
&\quad + \sqrt{n+1}\int\limits_{\SSS^2} d^2\vomega \, \lim_{r\searrow 0} \Bigl(\gamma +\delta\frac{\partial}{\partial r}  \Bigr) \Bigl( r \psi^{(n+1)}\bigl(y^n,r\vomega \bigr) \Bigr)\nonumber\\
&\quad +\: \frac{\hbar^2}{2m_y} \sum_{j=1}^n  \delta^3(\vy_j)
\int\limits_{\SSS^2} d^2\vomega \, \lim_{r\searrow 0} r\psi^{(n-1)}\bigl(y^n\setminus \vy_j,r\vomega\bigr)\,,
\label{Hdef1b}
\end{align}
where the constants $\gamma,\delta\in\CCC$ satisfy 
\begin{align}
&\alpha^* \gamma \in \RRR\label{ac}\\
&\beta^*\delta \in \RRR\label{bd}\\
\label{abcd}
&\alpha^*\delta-\gamma^*\beta=-1\,.
\end{align}
It should always be obvious when the symbol $\delta$ means the constant $\delta\in \CCC$ and when the Dirac delta function. Note that for $\gamma=0$, $\delta=g/4\pi$, the expression \eqref{Hdef1b} for the Hamiltonian agrees with \eqref{Hdef9a}, while for $\gamma=g/4\pi$, $\delta=0$ it reduces to \eqref{Hdef9b}. 

We remark that the family of IBCs just described actually depends only on 4 real parameters, although $\alpha,\beta,\gamma,\delta$ would seem at first to be 4 complex (and thus 8 real) parameters. That is so because \eqref{ac} requires $\alpha$ and $\gamma$ to have equal phases, and \eqref{bd} requires $\beta$ and $\delta$ to have equal phases; since the phase of $\alpha^* \delta$ is the phase difference between $\alpha$ and $\delta$, which must be equal to that between $\gamma$ and $\beta$ and thus by \eqref{abcd} to the phase of $-1$, $\alpha$ and $\delta$ must have equal phases. In other words, all of $\alpha,\beta,\gamma,\delta$ must have the same phase, while their absolute values are constrained by \eqref{abcd}, so that only 4 real parameters are independent.

On a non-rigorous level, the conservation of probability (i.e., self-adjointness of the Hamiltonian \eqref{Hdef1b}) can be checked by means of a calculation similar to the one in Section~\ref{sec:conservation} above (and to the one in 
\cite{co1}). 
The self-adjointness breaks down (already on the non-rigorous level) if we relax the conditions \eqref{ac}--\eqref{abcd}. If, however, \eqref{ac}--\eqref{abcd} are satisfied, then \eqref{IBC1b} and \eqref{Hdef1b} define a self-adjoint Hamiltonian, provided $E_0>0$ \cite{ibc2a}.

\item \emph{Dirac delta function terms in $H$.} \label{rem:Diracdelta} Here is a reason for thinking that, among the many different IBCs that are mathematically possible corresponding to different choices of the constants $\alpha,\beta,\gamma,\delta$, only the Dirichlet-type IBC \eqref{IBC9a}, corresponding to $\alpha=-4\pi/g$, $\beta=0=\gamma$, and $\delta=g/4\pi$, is physically relevant as a replacement of the original (UV divergent) Hamiltonian \eqref{Horigdef9}: It is the only choice that leads to a term in $H_{\mathrm{IBC}}$ that reproduces the Dirac delta function terms in $H_{\orig}$, i.e., for which the last line of \eqref{Hdef1b} agrees with the last line of \eqref{Horigdef9}. That is because the Dirichlet case is the only case in which the last line of \eqref{Hdef1b} can be expressed in terms of $\psi^{(n-1)}$.

And the last line of \eqref{Hdef1b} is dictated by the condition that $H\psi$ has to be an $L^2$ function and thus cannot contain contributions that are Dirac delta functions. Indeed, since for any $\alpha,\beta,\gamma,\delta$, the wave function $\psi^{(n)}$ diverges at the boundary like $1/r$, the Laplacian in the Hamiltonian always yields a distribution of the form 
\be\label{LaplacepsiDiracdelta}
-\sum_j\nabla^2_{\vy_j} \psi^{(n)}= -\sum_j \delta^3(\vy_j) \, f(y^n\setminus \vy_j) + g(y^n)
\ee
with some functions $f$ and $g$; so the Dirac delta contributions need to be canceled, which leads to the last line of \eqref{Hdef1b}.

In the 1-dimensional case, it seems that the physically reasonable IBC is the one given in \cite{KS15}, which is of the Neumann type and involves the normal derivative on both sides of the ``diagonal'' (i.e., the collision configurations). This is the conclusion one reaches when demanding that the Laplacian term in the Hamiltonian cancels the Dirac delta function terms, starting from the 1-dimensional version of \eqref{Horigdef9}: Then, the derivative of $\psi$ needs to have jumps of the appropriate magnitude.

\item \emph{Positivity.} Another reason for thinking that the Dirichlet-type IBC (rather than, say, Neumann-type) is the physically relevant choice may be that $H_{\mathrm{IBC}}$ for a Dirichlet-type IBC is positive, as mentioned in Theorem~\ref{thm:fixedx} in Section~\ref{sec:conservation}. Generally speaking, the IBC approach neither requires nor guarantees that Hamiltonians are bounded from below. 
Presently, we do not know for which other choices of $\alpha,\beta,\gamma,\delta$ the Hamiltonian will be positive, but we see reason to believe that, in the variant of the equations appropriate for Model 1, the Neumann-type IBC leads to a Hamiltonian that is not bounded from below. 
Further considerations about physical reasonableness of $H_{\mathrm{IBC}}$ can be found in Sections~\ref{sec:groundstate} and \ref{sec:removecutoff} below.

\item \emph{Bohmian trajectories.} There is a natural way of defining Bohmian trajectories for the models described in this paper; we describe this in detail elsewhere \cite{bohmibc}. The Bohmian configuration $Q_t$ follows a Markov jump process in configuration space that is $|\psi_t|^2$-distributed at every time $t$. The process has finitely many jumps in every finite time interval. The pieces between the jumps are solutions to Bohm's equation of motion; in particular, they are deterministic, in contrast for example to Nelson's trajectories \cite{Nel66}, which follow a diffusion process. In our Models 1 and 2, the jumps correspond to the creation or annihilation of a particle. The jumps to a lower sector (particle annihilation) occur whenever a $y$-particle hits an $x$-particle; in that event, the $y$-particle gets deleted from $Q_t$. While the jumps to a lower sector are deterministic, the jumps to a higher sector (particle creation) are stochastic. They can occur at any configuration $Q_t$, with a rate depending on $Q_t$ and $\psi_t$, and lead to a configuration with a new $y$-particle created at the location of an $x$-particle; the $y$-particle then moves in a random direction that is uniformly distributed over the sphere. Compared to previous models of particle creation and annihilation in Bohmian mechanics (see \cite{Tum04} and references therein) that involved a UV cut-off, the difference is that in Models 1 and 2, the $y$-particle gets created \emph{at} (rather than \emph{near}) an $x$-particle, and that annihilation is deterministic. The process is time-reversal invariant (notwithstanding that annihilation is deterministic and creation is stochastic).
\end{enumerate}

\subsection{IBC for Model 1}

We now describe an IBC and the corresponding Hamiltonian for Model 1. The IBC demands that for any $m,n\in\{1,2,\ldots\}$, any configuration $x^m=(\vx_1,\ldots,\vx_{m})\in \RRR^{3m}$ of $x$-particles, any configuration $y^n=(\vy_1,\ldots,\vy_{n})\in \RRR^{3n}$ of $y$-particles with $x^m\cap y^n = \emptyset$ (i.e., $\vx_i\neq \vy_j$ for all $i,j$), 
any $i=1,\ldots, m$, and any $j=1,\ldots, n$,
\be\label{IBC1a}
\lim_{(\vx_i,\vy_j)\to(\vx,\vx)} \,|\vy_j-\vx_i| \,\psi^{(m,n)}\bigl(x^m,y^n)
= - \frac{g}{2\pi\hbar^2\sqrt{n}} \frac{m_xm_y}{m_x+m_y}\, \psi^{(m,n-1)}(\vx_i=\vx,\widehat{\vy_j})\,,
\ee
where $\widehat{\ }$ denotes omission, 
and $g\in\RRR$ is the same coupling constant as before.

The IBC is a condition on the wave function $\psi$ near the \emph{diagonal} $\Delta$ in configuration space $\conf_x\times\conf_y$, i.e., the set of ``collision configurations,''
\be
\Delta = \Bigl\{ (x^m,y^n)\in\conf_x\times\conf_y: \vx_i=\vy_j \text{ for some }i,j \Bigr\}\,.
\ee
If we regard the collision configurations in $\Delta$ as not admissible configurations then the configuration space is the set difference $\conf=(\conf_x\times\conf_y)\setminus\Delta$, and its ``boundary'' is $\partial\conf=\Delta$. 

 The Hamiltonian $H_{\mathrm{IBC}}$ is defined \cite{Lam18} on a domain $\domain_{\mathrm{IBC}}\subset \Hilbert=\Fock^-\otimes\Fock^+$ consisting of wave functions $\psi$ that satisfy the IBC \eqref{IBC1a} and obey the following asymptotics, replacing \eqref{model2asymp}, near the boundary surface $\{\vx_i=\vy_j\}$ in $\Q^{(m,n)}= \RRR_x^{3m} \times \RRR_y^{3n}$:
\be\label{expand3}
\psi(x,y) = c_{-1,i}(x,y\setminus \vy_j) \, r_{ij}^{-1} + c_{\ell,i} (x,y\setminus \vy_j) \, \log r_{ij} + c_{0,i} (x,y\setminus \vy_j) + o(r_{ij}^0)\,,
\ee
where $r_{ij}=|\vx_i-\vy_j|$, $x=(\vx_1,\ldots,\vx_m)$, and $y=(\vy_1,\ldots,\vy_n)$ with $\vx_k\neq \vy_r$ for all $k,r$.
Moreover, in order to make $H_{\mathrm{IBC}}$ self-adjoint, the coefficients are related according to~\cite{Lam18}
\be\label{cell}
c_{\ell,i} = \eta\, c_{-1,i}
\ee
with fixed real proportionality factor
\be
\eta = \frac{g^2 \, m_x^2 m_y^2}{2\pi^2\hbar^4(m_x+m_y)^2} \Biggl[\frac{\sqrt{m_x(m_x+2m_y)}}{m_x+m_y}-\frac{m_x+m_y}{m_y}\arctan \biggl( \frac{m_y}{\sqrt{m_x(m_x+2m_y)}}  \biggr)  \Biggr]\,.
\ee
Note that as $m_x\to \infty$, the constant  $\alpha $ tends to zero, and by \eqref{cell}  the asymptotics \eqref{expand3}  reduces to  \eqref{model2asymp}, the one of Model~2.
In terms of the $c$ coefficients, the IBC \eqref{IBC1a} can be reformulated as 
\be\label{IBC1c}
c_{-1,i}(x,y\setminus \vy_j)= - \frac{g}{2\pi\hbar^2\sqrt{n}} \frac{m_xm_y}{m_x+m_y}\,\psi(x,y\setminus \vy_j )  \,.
\ee

The Hamiltonian $H_{\mathrm{IBC}}$ is given by
\begin{align}
(H_{\mathrm{IBC}}\psi)^{(m,n)}(x^m,y^n) 
&= -\frac{\hbar^2}{2m_x} \sum_{i=1}^{m} \nabla^2_{\vx_i}\psi^{(m,n)}(x^m,y^n)
-\frac{\hbar^2}{2m_y} \sum_{j=1}^{n} \nabla^2_{\vy_j}\psi^{(m,n)}(x^m,y^n)\nonumber\\[2mm]
&\quad+nE_0 \psi^{(m,n)}(x^m,y^n) 
+g\sqrt{n+1}\sum_{i=1}^{m} c_{0,i}(x^m,y^n) \nonumber\\[2mm]
&\quad+ \frac{g}{\sqrt{n}} \sum_{i=1}^m \sum_{j=1}^n  \delta^3(\vx_i-\vy_j)\,\psi^{(m,n-1)}\bigl(x^m,y^n\setminus \vy_j\bigr)\,.\label{Hdef1a}
\end{align}
As shown in \cite{Lam18}, $H_{\mathrm{IBC}}$  defines a self-adjoint operator on $\domain_{\mathrm{IBC}}$.
Of the five terms in \eqref{Hdef1a}, only the $nE_0 \psi$ term is square integrable by itself. All the other terms are distributions that add up to a square integrable function. In particular, in contrast to the Hamiltonian \eqref{Hdef9a} of Model~2, the annihilation term $c_{0,i}(x^m,y^n)$ is not necessarily square integrable but compensates derivatives of the logarithmic contributions in \eqref{expand3}. These cancelations are rather subtle, which could be a reason why no renormalization scheme for Model~1 was known before the IBC study in \cite{Lam18}.

\subsection{Ground State Energy and Effective Yukawa Potential}
\label{sec:groundstate}

Elsewhere \cite{ibc2a}, we show that $H_{\mathrm{IBC}}$ as in \eqref{Hdef9a} for Model 2 with Dirichlet-type IBC \eqref{IBC9a} and $E_0>0$ possesses a non-degenerate ground state $\psi_{\min}$. It is given by
\be
\psi_{\min}(\vy_1,\ldots,\vy_n) =\mathcal{N}\frac{1}{\sqrt{n!}} \Bigl( -\frac{gm_y}{2\pi\hbar^2}\Bigr)^{\! n} \prod_{j=1}^n\frac{e^{-\sqrt{2m_yE_0}|\vy_j|/\hbar}}{|\vy_j|} 
\ee
with normalization constant
\be
\mathcal{N} = \exp \Bigl(-\frac{g^2m_y^2}{4\pi\hbar^3\sqrt{2m_yE_0}}  \Bigr)
\ee
and eigenvalue
\be\label{Emin}
E_{\min}=\frac{g^2m_y\sqrt{2m_yE_0}}{2\pi\hbar^3}\,.
\ee
That is, the state is a superposition of different numbers of $y$-particles, and in each sector all $y$-particles have the same wave function; so the $x$-particle at the origin is dressed with a cloud of $y$-particles. The probability distribution of the number $n$ of $y$-particles is a Poisson distribution with mean value
\be
\langle n \rangle_{\psi_{\min}} = \frac{g^2 m_y^2}{2\pi\hbar^3\sqrt{2m_yE_0}}\,.
\ee

We claim further that in Model 1, in which the $x$-particles interact by exchanging $y$-particles, the $x$-particles effectively interact through a Yukawa potential; this in fact agrees with the result of Yukawa's original reasoning \cite{Yuk35}, see below. A simple way of computing the effective interaction potential is to consider (as ``Model 2b'') $N$ $x$-particles fixed at $\vx_1,\ldots,\vx_N\in\RRR^3$ and to find the ground state energy. To this end, consider wave functions of $y$-configurations that simultaneously satisfy $N$ IBCs,
\be\label{IBC9d}
\lim_{r\searrow 0} \biggl(r\psi^{(n+1)}(y^n,\vx_i+r\vomega)  \biggr) 
= - \frac{g\, m_y}{2\pi\hbar^2\sqrt{n+1}}\psi^{(n)}(y^n)
\ee
for every $i=1,\ldots,N$, $\vomega\in\SSS^2$, $n\in\{0,1,2,\ldots\}$, and $y^n\in(\RRR^3\setminus\{\vx_1,\ldots,\vx_N\})^n$. The corresponding Hamiltonian reads
\begin{align}
(H\psi)^{(n)}(y^n) 
&= -\frac{\hbar^2}{2m_y} \sum_{j=1}^{n} \nabla^2_{\vy_j}\psi^{(n)}(y^n)+ nE_0 \psi^{(n)}(y^n) \nonumber\\[2mm]
&\quad +\: \frac{g\sqrt{n+1}}{4\pi}\sum_{i=1}^N \int\limits_{\SSS^2} d^2\vomega \, \lim_{r\searrow 0} \frac{\partial}{\partial r} \Bigl( r \psi^{(n+1)}(y^n,\vx_i+r\vomega) \Bigr)\nonumber\\
&\quad +\: \frac{g}{\sqrt{n}} \sum_{i=1}^N\sum_{j=1}^n  \delta^3(\vy_j-\vx_i)\,\psi^{(n-1)}(y^n\setminus \vy_j)\,.\label{Hdef9d}
\end{align}
For $E_0>0$, the ground state is
\be
\psi_{\min}(\vy_1,\ldots,\vy_n)=c_n \prod_{j=1}^n \sum_{i=1}^N \frac{e^{-\sqrt{2m_yE_0}|\vy_j-\vx_i|/\hbar}}{|\vy_j-\vx_i|} 
\ee
with suitable factors $c_n$ and eigenvalue
\be\label{Yukawa}
E_{\min}=\frac{g^2m_y}{\pi\hbar^2}\biggl( \frac{N\sqrt{2m_yE_0}}{2\hbar}-\sum_{1\leq i<j\leq N}\frac{e^{-\sqrt{2m_yE_0}|\vx_i-\vx_j|/\hbar}}{|\vx_i-\vx_j|} \biggr)\,.
\ee
That is, the ground state energy of the $y$-particles, given the $x$-particles at $\vx_1,\ldots,\vx_N$, is given by \eqref{Yukawa}. Regarding this energy function of $\vx_1,\ldots,\vx_N$ as an effective potential for the $x$-particles (which is appropriate when the $x$-particles move slowly), we see that $x$-particles effectively interact through an attractive Yukawa pair potential,
\be
V(R)=\text{const.}-\frac{e^{-\lambda R}}{R}
\ee
with $R$ the distance between two $x$-particles. If we take the energy needed to create a $y$-particle to be $E_0=m_yc^2$, then
\be
\lambda=\sqrt{2}\frac{m_y c}{\hbar}\,, 
\ee
which is, up to the factor $\sqrt{2}$, the value originally obtained by Yukawa \cite{Yuk35} considering the effective interaction of nucleons by exchange of pions. We expect that the factor $\sqrt{2}$ is owed to the non-relativistic nature of our model.

\subsection{IBC Hamiltonians as a Limit of Removing the Cut-Off}
\label{sec:removecutoff}

If one introduces a UV cut-off into the UV divergent original Hamiltonian \eqref{Horigdef9} of Model 2, it becomes the well-defined operator $H_\varphi$ given by
\begin{align}
(H_\varphi \psi)^{(n)}(y^n) &= -\frac{\hbar^2}{2m_y} \sum_{j=1}^n \nabla_{\vy_j}^2 \psi^{(n)}(y^n)+ nE_0 \psi^{(n)}(y^n) \nonumber\\
&\quad +\: g \sqrt{n+1} \int_{\RRR^3} d^3\vy\, \varphi^*(\vy)\, \psi^{(n+1)}(y^n,\vy) \nonumber\\
&\quad +\: \frac{g}{\sqrt{n}} \sum_{j=1}^n \varphi(\vy_j) \,\psi^{(n-1)}(y^n\setminus \vy_j) \,,\label{Hcutoffdef9}
\end{align}
analogous to \eqref{Hcutoffdef} for Model 1. Here, the Dirac delta function $\delta^3$ has been replaced by $\varphi:\RRR^3\to\CCC$, a square-integrable function describing the charge density of the $x$-particle, and the limit $\varphi\to\delta^3$ would correspond to removing the UV cut-off. It is well known \cite{vH52,Der03} that, if $E_0>0$, there are numbers $E_\varphi\in\RRR$ such that the operator $H_\varphi - E_\varphi$ possesses a limit as $\varphi\to \delta^3$. Since $E_\varphi\to\infty$ in this limit, $H_\varphi$ tends to infinity in a sense, but this sense is harmless because two Hamiltonians that differ only by a multiple of the identity operator can be regarded as equivalent, as they generate the same time evolution (if we regard wave functions that differ only by a global phase factor as equivalent). So, in a relevant sense, the limit $\varphi\to\delta^3$ can indeed be taken,
which suggests regarding the limiting Hamiltonian as the ``physically correct'' Hamiltonian.
Obviously, the limiting Hamiltonian is defined only up to addition of a constant, as this constant can be added to each of the $E_\varphi$. However, for the $H_{\varphi}$ of this concrete model, there is a particular natural choice of $E_\varphi$ and thus of the limiting Hamiltonian, described, e.g., in \cite{Der03,ibc2a}, for which we will write $H_\infty$.

\begin{thm}\label{thm:renormalized} {\rm \cite{ibc2a}}
For the Hamiltonian $H_{\mathrm{IBC}}$ of the Dirichlet type for Model 2, defined by \eqref{IBC9a} and \eqref{Hdef9a}, 
\be
H_\infty = H_{\mathrm{IBC}}-E_{\min}
\ee
with the constant $E_{\min}$ as in \eqref{Emin}.
\end{thm}

On the one hand, this result lends further support to regarding $H_{\mathrm{IBC}}$ and the IBC approach as physically reasonable. On the other hand, the result provides a more direct and explicit representation of $H_\infty$ than was available so far.

\section{Conclusions}
\label{sec:conclusions}

If configuration space has a boundary (be it of codimension 1 or codimension 3), then an interior--boundary condition on this boundary can serve to ensure that the probability flux into the boundary is compensated by an equally large gain of probability in another place in configuration space, and thus to ensure the overall conservation of probability. At the same time, an interior--boundary condition has the consequence that contributions to the wave function can flow out of the boundary. That is just what is needed for defining an evolution on Fock space representing the emission and absorption of particles. The crucial point here is that this approach is also possible if the sources (the $x$-particles in Model 1 and Model 2) are point-shaped, without a UV divergence problem arising. IBCs thus seem like a natural way of implementing Hamiltonians with particle creation and annihilation. The mathematical viability of this approach for non-relativistic QFTs has been established in recent works \cite{ibc2a,LS18,Lam18}. Its applicability to more serious QFTs remains to be explored.

\bigskip

\noindent\textit{Conflict of interest.} On behalf of all authors, the corresponding author states that there is no conflict of interest. 

\bigskip

\noindent\textit{Acknowledgments.} 
We are grateful to Sheldon Goldstein, Stefan Keppeler, Jonas Lampart, and Julian Schmidt for helpful discussions.

\end{document}